\documentclass{elsarticle}

\newtheorem{definition}{\textbf{Definition}}

\begin{document}

\begin{frontmatter}

\title{\textbf{A Practical Runtime Security Policy Transformation Framework for Software Defined Networks}}

\author[1]{Yunfei Meng}
\author[2]{Changbo Ke}
\author[3]{Zhiqiu Huang}
\author[3]{Guohua Shen}
\author[1]{ChunQiang Liu}
\author[1]{Xiaojie Feng}
\address[1]{College of Information Engineering, Qingdao Binhai University, Qingdao 266555, China}
\address[2]{School of Computer Science and Technology, Nanjing University of Posts and Telecommunications, Nanjing 210023, China}
\address[3]{College of Computer Science and Technology, Nanjing University of Aeronautics and Astronautics, Nanjing 211106, China}

\begin{abstract}
Software-defined networking (SDN) has been widely utilized to enforce the security of traditional networks, thereby promoting the process of transforming traditional networks into SDN networks. However, SDN-based security enforcement mechanisms rely heavily on the security policies containing the underlying information of data plane, such as MAC address, IP address or switch ports. These security policies need to be specifically developed by the network operators, and loaded into the control plane by manual inputting. With increasing the scale of underlying network, the current security policy management mechanism will confront more and more challenges. The security policy transformation for SDN networks is to research how to transform the high-level security policy without containing the underlying information of data plane into the practical flow entries used by the OpenFlow switches automatically, thereby implementing the automation of security policy management. Based on this insight, a practical runtime security policy transformation framework is proposed in this paper. First of all, we specify the security policies used by SDN networks as a system model of security policy (SPM). From the theoretical level, we establish the system model for SDN network and propose a formal method to transform SPM into the system model of flow entries automatically. From the practical level, we propose a runtime security policy transformation framework to solve the problem of how to find a connected path for each relationship of SPM in the data plane, as well as how to generate the practical flow entries according to the system model of flow entries. In order to validate the feasibility and effectiveness of the framework, we set up an experimental system and implement the framework with POX controller and Mininet emulator. The experimental results illustrate the framework can synchronously perceive the changes caused by cutting down one edge or changing SPM, and keep the data plane holding the security properties defined by SPM continuously at runtime.
\end{abstract}


\end{frontmatter}

\section{Introduction}
Software-defined networking (SDN) is a novel networking technique or architecture that changes the limitation of traditional network infrastructures by breaking the vertical integration, decoupling the control logics from the underlying forwarding devices, promoting the centralization of control and introducing the abilities to program the network directly\cite{SDN2}. In SDN networks, the control logics of network, such as routing, traffic engineering or security policy developed in the application plane, are loaded into the control plane via the northbound interfaces (NBI) and transformed into a set of forward entries used by the OpenFlow switches. After that, the control plane distributes the generated flow entries to the associated switches in the data plane via the southbound interfaces (SBI). Because of its programmable, centralized intelligent control as well as global traffic view, SDN has been widely utilized to enhance the security of tradition networks, thereby promoting the process of transforming traditional networks into SDN networks. For instances, Garay et al.\cite{66} proposed a SDN-based network access control mechanism, flownac, which is a centralized EAP (extensible authentication protocol) for IEEE 802.1x wireless local area network (WLAN). Yakasai et al.\cite{67} proposed a network access control mechanism, flowidentity. This mechanism integrates EAP security authentication mechanism into the SDN controller. Hu et al.\cite{68} proposed a dynamic firewall mechanism, flowguard, based on SDN. Koerner et al.\cite{69} proposed a device security authentication mechanism based on MAC address and SDN.

However, SDN-based security enforcement mechanisms rely heavily on the security policies containing the underlying information of the data plane, such as MAC address, IP address or switch ports. These security policies need to be specifically developed by the network operators, and loaded into the control plane by means of the manual inputting. With increasing the scale of underlying network, the current security policy management mechanism will confront more and more challenges. First of all, it is nearly impossible for any operator to completely understand all the information of underlying network. In addition, with the emergence of multi-controller SDN\cite{70}, network operators need to manage a variety of heterogeneous controllers at the same time. In this case, the same security policy often needs to be developed and deployed for the different types of controller, which inevitably increases the complexity and difficulty for network management. Therefore, a novel security policy management mechanism which can be completely transparent to the underlying information of data plane is  urgently needed for SDN networks. That is, it can permit the operators only to define the high-level security policy without containing any underlying information, then by means of the security policy transformation, the high-level security policy can be automatically transformed into its corresponding flow entries used by the OpenFlow switches in the data plane, thereby implementing the automation of security policy management for SDN networks.

Based on these insights, we have proposed a security policy model transformation and verification approach for SDN networks and published the approach in our previous paper\cite{meng}. In that paper, we proposed a security policy transformation method to transform the high-level security policy model (SPM) without containing the underlying information into its corresponding low-level security policy model (LSPM) containing the underlying information. To verify the soundness of proposed security policy model transformation method, we further proposed a security policy verification method and proved that the problem of whether the data plane can satisfy the security properties defined by SPM is equivalent to the problem of searching the connected paths related with SPM in the data plane, that is, as long as each access control relationship $R_i\in$ SPM can be transformed into a corresponding connected path $P_i$ in the data plane, next transforms $P_i$ into a set of flow entries used by the OpenFlow switches, then the data plane must can hold the security properties defined by SPM. However, that paper only proposed the method from the theoretical level, and did not specifically implement this method. Moreover, it did not solve the problem of how to find a connected path for each relationship of SPM in the data plane, and how to transform LSPM into the practical flow entries used by the switches.

And based on the theoretical foundation of that paper, we propose a runtime security policy transformation framework for SDN networks in this paper. First of all, this paper further improves the system model of SDN networks and solves the problem of how to transform SPM into the flow entries used by the OpenFlow switches from the theoretical level. Moreover, this paper proposes a runtime security policy transformation framework from the practical level, thereby solving the problem of how to find a connected path $P_i$ for each relationship $R_i\in$ SPM in the data plane, as well as how to transform the system model of flow entries into the practical flow entries used by the switches at runtime. In addition, this paper further implements the proposed framework with an experimental system. The experimental result illustrate the framework is completely effective at runtime.

Hence, the contributions of this paper can be summarized as follows:

$\bullet$ We specify the security policies used by SDN networks, such as access control policies or firewall policies, as a system model of security policy (SPM). SPM is of a high-level system model without containing any underlying information of data plane.

$\bullet$ From the theoretical level, we establish the system model for SDN network, and propose a formal method to transform SPM into the system model of flow entries automatically. The system model of flow entry is of a low-level system model containing the underlying information of data plane.

$\bullet$ From the practical level, we propose a runtime security policy transformation framework which consists of the security policy module, topology discovery module, runtime monitoring module, path generation module and flow entry generation module. Leveraging these functional modules, the framework can solve the problem of how to find a connected path for each relationship of SPM in the data plane, how to transform the path into the system model of flow entries, as well we how to generate the practical flow entries by using the system model of flow entries.

$\bullet$ In order to validate the feasibility and effectiveness of the framework, we set up an experimental system and implement the framework by using POX controller and Mininet emulator. The experimental result illustrate the framework is completely effective at runtime.

The remainder of this paper is structured as follows. Section 2 discusses some related works. Section 3 proposes the system model and elaborates on how to transform SPM into the flow entries from the theoretical level. Section 4 proposes the runtime security policy transformation framework from the practical level and introduces its functional modules. Section 5 implements the framework with an experimental system and elaborates on how to evaluate the effectiveness and performance of the framework. Finally, Section 6 concludes this paper and presents some future directions.

\section{Related Work}
In this section, we discuss some research works related with the policy model transformation and the security policy verification.

\subsection{Policy Model Transformation}
According to the definitions of model-driven architecture (MDA), the model transformation refers to the process of transforming the platform independent model (PIM) to its corresponding platform specific model (PSM)\cite{71}. As far as the literatures we have read, the researches towards the policy model transformation can be roughly divided into three categories, they are the template-based transformation, RBAC-oriented transformation as well as the transformation based on the system model and mapping rules\cite{72}. Due to the limitation of template, the template-based model transformation has very limited transformation capability. Generally, RBAC-oriented model transformation\cite{74} is only suitable for transforming RBAC (role-based access control) policies, and does not have enough capability to describe the complex system, so that these two methods are not suitable for SDN networks.

At present, the model transformation based on the system model and mapping rules has been widely used for transforming the policy models. The main idea of this method can be summarized as follows: (1) \emph{System Model}: it defines the objects of system and the relationship between the system objects; (2) \emph{Policy Model}: it defines the policy object and the relationship between the policy objects; (3) \emph{Mapping Rules}: it establishes the mapping rules between the upper-level policy objects and the lower-level system objects\cite{75}\cite{76}. The transformation based on the system model and mapping rules first establishes the policy model and the system model which can describe the underlying system, then establishes the mapping rules between the policy objects and the system objects, then transforms the upper-level policy model into its corresponding lower-level policy model by means of the established mapping rules. In particular, Davy et al.\cite{77} proposed a policy model transformation method based on mapping rules, in which the policy model is defined as a tuple ( event, condition, behavior, subject, object ) and used the ontology to establish the mapping rules between the different system layers. Luck et al.\cite{78} proposed a method to transform RBAC model defined in service layer into the policy model used in the system layer. In this method, the system model is divided into three layers: roles and object (RO), subject and resources (SR) and processes and hosts (PH), and the mapping rules between the three layers have been established. Based on the Luck's research, Porto et al.\cite{79} further decomposes the PH layer into two sub layers, namely DAS (diagram abstract subsystem) layer and PH layer. DAS layer is mainly used to describe the network topology in the original PH layer, while PH layer is used to describe the specific network information in DAS layer. In addition, the authors also proposed a policy verification framework, which can be used to verify the consistency problems in the process of policy transformation. In addition, Lampson et al.\cite{80} proposed a network policy model transformation method for the distributed computing environment. Maullo et al.\cite{81} proposed a policy transformation system based on the first-order predicate logics, which transforms the high-level policy model into the low-level network configuration policy through the network topology and other information. Nanxi et al.\cite{82} proposed a SDN-oriented access control policy transformation framework. In this paper, In this paper, we also propose a security policy transformation framework based on the system model and mapping rules. We first establish the system model of security policy (SPM) and data plane, then establish the transformation rules between the policy objects of SPM and the objects of the data plane, thereby transforming SPM into the system model of flow entries automatically.

\subsection{Security Policy Verification}
To assure the information systems running securely, security mechanisms of information system need to be validated whether it can satisfy the security properties defined by the security policy. The traditional validation methods based on the testing and simulation can only confirm the system can work properly under the different testing scenarios, but it is difficult to find some hidden scenarios that occur with little probability. Formal verification methods have been applied to overcome the shortcomings existed in the traditional validation methods. At present, the formal verification methods for validating the security policy mainly include theorem proving and model checking\cite{Ma}. Theorem proving is unsuitable to validate the properties of complex systems due to its lower efficiency. Model checking\cite{Clarke} can be used to validate whether the system model can satisfy the expected dynamic behaviors and specific static properties. Model checking technique has been widely used for the security policy verification. For instances, Al-Shaer et al.\cite{Shaer} proposed a static policy inconsistency detection method for the firewall policies of network. Bandara et al.\cite{Bandara} proposed a security policy verification framework based on event calculus (EC) and used the reasoning techniques for the policy conflict identification. May et al.\cite{May} verified the privacy policies by means of an asynchronous model checker. Rubio-Loyola et al.\cite{Rubio} proposed a goal-oriented policy refinement and conflict detection method by means of the model checking technique and linear temporal Logic (LTL). Graham et al.\cite{Graham} proposed a policy conflict detection method with the model checking and an extended decision table. Baliosan and Serrat\cite{Baliosian} proposed a specific finite automata based method for the policy conflict detection.

\section{Problem Formalization}
The security policy transformation for SDN networks is to research how to transform the high-level security policy without containing the underlying information into the set of practical flow entries used by the OpenFlow switches in the data plane automatically, thereby implementing the automation of security policy management in SDN network. In the following of this section, we first establish the system model for SDN network, then propose a formal method to transform the security policy (SPM) into the system model of flow entries from the theoretical level.

\subsection{ System Model}
\begin{definition}(Security Policy):
\emph{
The high-level security policy is defined as a finite set of access control relationships: SPM = $\{$ $R_{0}$, $R_{1}$,...,$R_{n}$ $|\ \forall$ $R_{i}$ = $($ $s_i$, $o_j$, $a$ $)$ $\}$, where $s_i\in\mathcal{S}$ represents the subject of the relationship, $o_j\in\mathcal{O}$ represents the object of the relationship, $a$ represents the access authorization, i.e., the subject can access the object.
}
\end{definition}

\begin{definition}(Host):
\emph{
The host existed in the data plane is defined as a tuple: $h_i$=$($ $ip_i$, $sw_i$, $port_{i}^{m}$ $)$, where $ip_i$ represents the host's IP address in the data plane, $sw_i$ represents the OpenFlow switch connected with the host, $port_{sw_i}^{m}$ represents the port connected with the host in $sw_i$.
}
\end{definition}

\begin{definition}(OpenFlow Switch):
\emph{
The OpenFlow switch existed the data plane is defined as a finite set of flow entries: $sw_i$ = $\{$ $f_0$, $f_1$, ..., $f_n$ $\}$.
}
\end{definition}

\begin{definition}(Flow Entry):
\emph{
The flow entry existed in the OpenFlow switch is defined as a tuple: $f_{i}$ = $($ $ip_{src}$, $ip_{dst}$, $port_{sw_i}^{in}\Longrightarrow port_{sw_i}^{out}$ $)$, where $ip_{src}$ represents the traffic's source IP address, $ip_{dst}$ represents the traffic's destination IP address, $port_{sw_i}^{in}\Longrightarrow port_{sw_i}^{out}$ represents the traffic input from $port_{sw_i}^{in}$ will be outputted from $port_{sw_i}^{out}$ in the switch $sw_i$.
}
\end{definition}

\begin{definition}(Edge):
\emph{
The edge existed in the data plane is defined as: $e_i$ = $port_{start}^{out}\longmapsto port_{end}^{in}$, where $port_{start}^{out}$ represents the port connected with the edge in the start switch $sw_{start}$, $port_{end}^{in}$ represents the port connected with the edge in the end switch $sw_{end}$, so that the direction of the edge is from $sw_{start}$ to $sw_{end}$.
}
\end{definition}

\begin{definition}(Topology):
\emph{
The topology of the data plane is defined as an graph: $\mathcal{G}$ = $($ $SW$, $E$ $)$, where $SW$ represents a finite set of OpenFlow switches, $E$ represents a finite set of edges.
}
\end{definition}

\begin{definition}(Connected Path):
\emph{
The connected path between the host $h_i$ and the host $h_j$ in the topology $\mathcal{G}$ is defined as: $P_i$ = $h_i\longmapsto h_j$ = $\{$ $h_i$, $e_0$, $e_1$,..., $e_n$, $h_j$ $|$ $\forall e_i\in E$ $\}$, where the direction of the path is from $h_i$ to $h_j$.
}
\end{definition}

\subsection{Transforming SPM into the Flow Entries }
Based on the established system model, we propose a formal method to transform SPM into the system model of flow entries from the theoretical level. The method can be described as Figure 1 and summarized as follows: First of all, for $\forall R_i\in$ SPM, the subject $s_i\in R_i$ is transformed into a host $h(s_i)$ in the data plane, the object $o_j\in R_i$ is transformed into a host $h(o_i)$ in the data plane; Next, the access authorization $a\in R_i$ is transformed into a connected path $P_i$ between $h(s_i)$ and $h(o_j)$; After that, the connected path $P_i$ is transformed into a set of flow entries used by the switches which are passed by $P_i$; Finally, SPM is transformed into the set of flow entries $\Delta$ when all relationships of SPM have been transformed. As shown in Figure 1, the relationship $R_1\in$ SPM is transformed into the connected path $P_1$ between $h_1$ and $h_2$, thus $P_1$ is transformed into the flow entries deployed in $sw_1$, so as to implement the security policy transformation from the system model level. The soundness of the formal method has been proven in our previous paper\cite{meng}. That is, if the security properties defined by SPM is denoted as $\varphi$, the system model of data plane is denoted as $\mathcal{D}$, the flow entries generated by the method is denoted as $\Delta$, then the method can ensure the data plane $\mathcal{D}$ loaded with $\Delta$ can synchronously and continuously hold the security properties $\varphi$ at runtime, i.e., $\mathcal{D}(\Delta)$ $\models$ $\varphi$.
\begin{figure}[htbp]
\centering
\scalebox{0.30}{\includegraphics{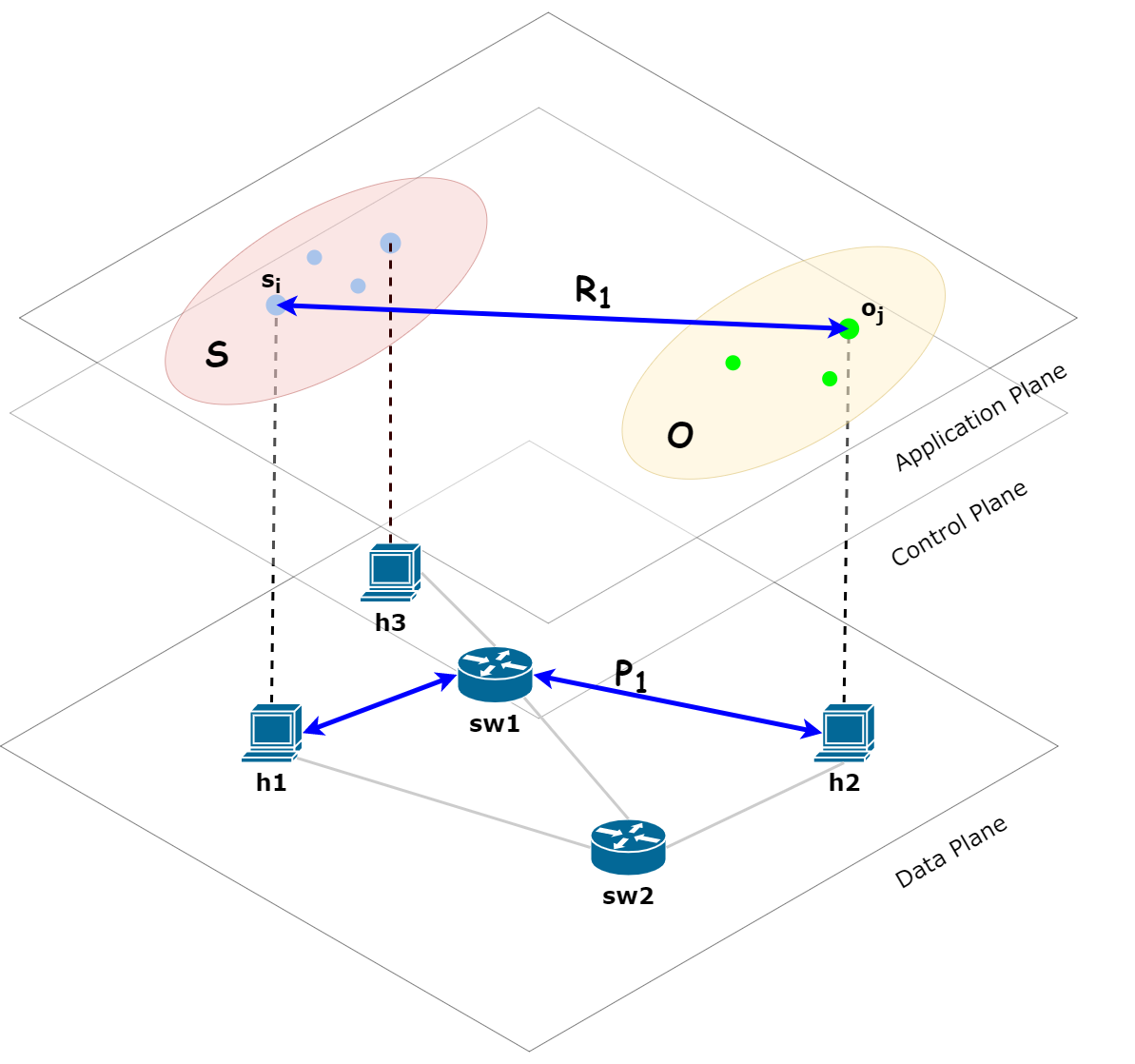}}
\caption{ The method of transforming SPM into the system model of flow entries. }
\end{figure}

Specifically, the formal method to transform SPM into the system model of flow entries is defined as follows.

$\bullet$ \textbf{ Transforming the Subject} Given an access control relationship $R_{i}\in$ SPM and the subject $s_{i}\in R_i$, then $s_{i}$ is transformed into a corresponding host existed in the data plane. The rule of transforming the subject of SPM is formally defined as follows:
\begin{equation}
\frac{s_i\in\ R_i}{h(s_{i})=( ip_{src},\ sw_{src},\ port_{src}^{in},\ )} \\
\end{equation}
where $h(s_i)$ represents the host mapped with $s_i$ in the data plane, $ip_{src}$ represents the IP address of $h(s_i)$ in the data plane, $sw_{src}$ represents the OpenFlow switch connected with $h(s_i)$ in the data plane, $port_{src}^{in}$ represents the port connected with $h(s_i)$ in $sw_{src}$.

$\bullet$ \textbf{ Transforming the Object} Given an access control relationship $R_{i}\in$ SPM and the object $o_{j}\in R_i$, then $o_{j}$ is transformed into a corresponding host existed in the data plane. The rule of transforming the object of SPM is formally defined as follows:
\begin{equation}
\frac{o_j\in\ R_i}{h(o_{j})=( ip_{dst},\ sw_{dst},\ port_{dst}^{out},\ )} \\
\end{equation}
where $h(o_j)$ represents the host mapped with $o_j$ in the data plane, $ip_{dst}$ represents the IP address of $h(o_j)$ in the data plane, $sw_{dst}$ represents the OpenFlow switch connected with $h(o_j)$, $port_{dst}^{out}$ represents the port connected with $h(o_j)$ in $sw_{dst}$.

$\bullet$ \textbf{ Transforming the Authorization} Given an access control relationship $R_{i}\in$ SPM and the access authorization $a\in R_i$, if there existing a connected path $P_i$ between $h(s_i)$ and $h(o_j)$ in the topology, then $a$ is transformed into the connected $P_i$. The rule of transforming the authorization is formally defined as follows:
\begin{equation}
\frac{a\in R_i}{h(s_i)\longmapsto h(o_j)\subset\mathcal{G}} \\
\end{equation}
where $h(s_i)\longmapsto h(o_j)$ represents a directional connected path from $h(s_i)$ to $h(o_j)$ in the topology.

$\bullet$ \textbf{ Transforming the Path} The connected path $h(s_i)\longmapsto h(o_j)$ is transformed into a set of flow entries deployed in the OpenFlow switches which are passed by the path. The rule of transforming the connected path is formally defined as follows:
\begin{equation}
\frac{h(s_i)\longmapsto h(o_j)}{\bigcup\limits_{k=src}^{dst}f_{i}(k)} \\
\end{equation}
where $f_{i}(k)$ =$($ $ip_{src}$, $ip_{dst}$, $port_{k}^{in}\Longrightarrow port_{k}^{out}$ $)$ represents a flow entry deployed in the switch $sw_k$ which is passed by $h(s_i)\longmapsto h(o_j)$. Leveraging the definitions of the system model, we can proof that the rule of transforming the connected path is sound.
\begin{displaymath}
\begin{array}{l}
proof:\\
\ \ \ h(s_i)\longmapsto h(o_j)\\
=\{\ h(s_i),\ e_0,\ e_1,\ ...,\ e_n,\ h(o_j)\ \}\\
=\{ (ip_{src},\ sw_{src},\ port_{src}^{in}),\ (port_{src}^{out}\longmapsto port_{0}^{in})\\
\ \ \ ,...,(port_{n}^{out}\longmapsto port_{dst}^{in}),\ (ip_{dst},\ sw_{dst},\ port_{dst}^{out}) \}\\
=\{ (port_{src}^{in}\Longrightarrow port_{src}^{out}),\ (port_{0}^{in}\Longrightarrow port_{0}^{out}),\\
\ \ \ ,...,(port_{n}^{in}\Longrightarrow port_{n}^{out}),\ (port_{dst}^{in}\Longrightarrow port_{dst}^{out})\\
\ \ (ip_{src},\ ip_{dst}),\ (sw_{src},\ sw_{dst}) \}\\
=\{ \bigcup\limits_{k=src}^{dst} ( port_{k}^{in}\Longrightarrow port_{k}^{out} ) \}\bigcup\ (ip_{src},\ ip_{dst}) \\
=\bigcup\limits_{k=src}^{dst} \{ ( port_{k}^{in}\Longrightarrow port_{k}^{out} ) \bigcup\ (ip_{src},\ ip_{dst}) \}\\
=\bigcup\limits_{k=src}^{dst}f_{i}(k).\\
\end{array}
\end{displaymath}

Therefore, $\forall R_i\in$ SPM, the subject $s_i\in R_i$, the object $o_j\in R_i$ and the autherization $a\in R_i$,
if $\exists P_i$ = $h(s_i)\longmapsto h(o_j)$ in the topology, then $R_i$ can be transformed into a corresponding set of flow entries $\bigcup\limits_{k=src}^{dst}f_{i}(k)$ by using the system model step by step, it can be formally defined as follows:
\begin{equation}
\frac{R_i}{\bigcup\limits_{k=src}^{dst}f_{i}(k)} \\
\end{equation}

$\bullet$ \textbf{ Transforming SPM} SPM is transformed into a corresponding set of flow entries $\Delta$ by using the equation (5). If $||$ SPM $||$ = $m$, then the rule of transforming SPM is formally defined as follows:
\begin{equation}
\Delta=\bigcup\limits_{i=1}^{m}\{ \bigcup\limits_{k=src}^{dst}f_{i}(k) \}
\end{equation}

\section{The Security Policy Transformation Framework}
The problem of how to transform SPM into the corresponding flow entries used by the OpenFlow switches has been solved from the theoretical level in Section 3. However, we cannot solve the problem of how to find a connected path in the data plane for each relationship $R_i\in$ SPM. In addition, by means of the security policy transformation method, SPM can be transformed into the corresponding flow entries, but the flow entry transformed from SPM is of the system model, i.e. it is only the formal description of the real flow entry and cannot be used by the real OpenFlow switch directly,  so that we need to further solve the problem of how to generate the practical flow entries by using the system model of flow entries. Based on this insight, a runtime security policy transformation framework for SDN networks is proposed from the practical level in this section. By means of the framework, we can solve the problem of how to find the connected path for each relationship defined by SPM, as well as the problem of how to generate the practical flow entries based on the system model at runtime. As shown in Figure 2, this framework consists of 5 functional modules, i.e., the security policy module, the topology discovery module, the runtime monitoring module, the path generation module and the flow entry generation module.
\begin{figure}[htbp]
\centering
\scalebox{0.22}{\includegraphics{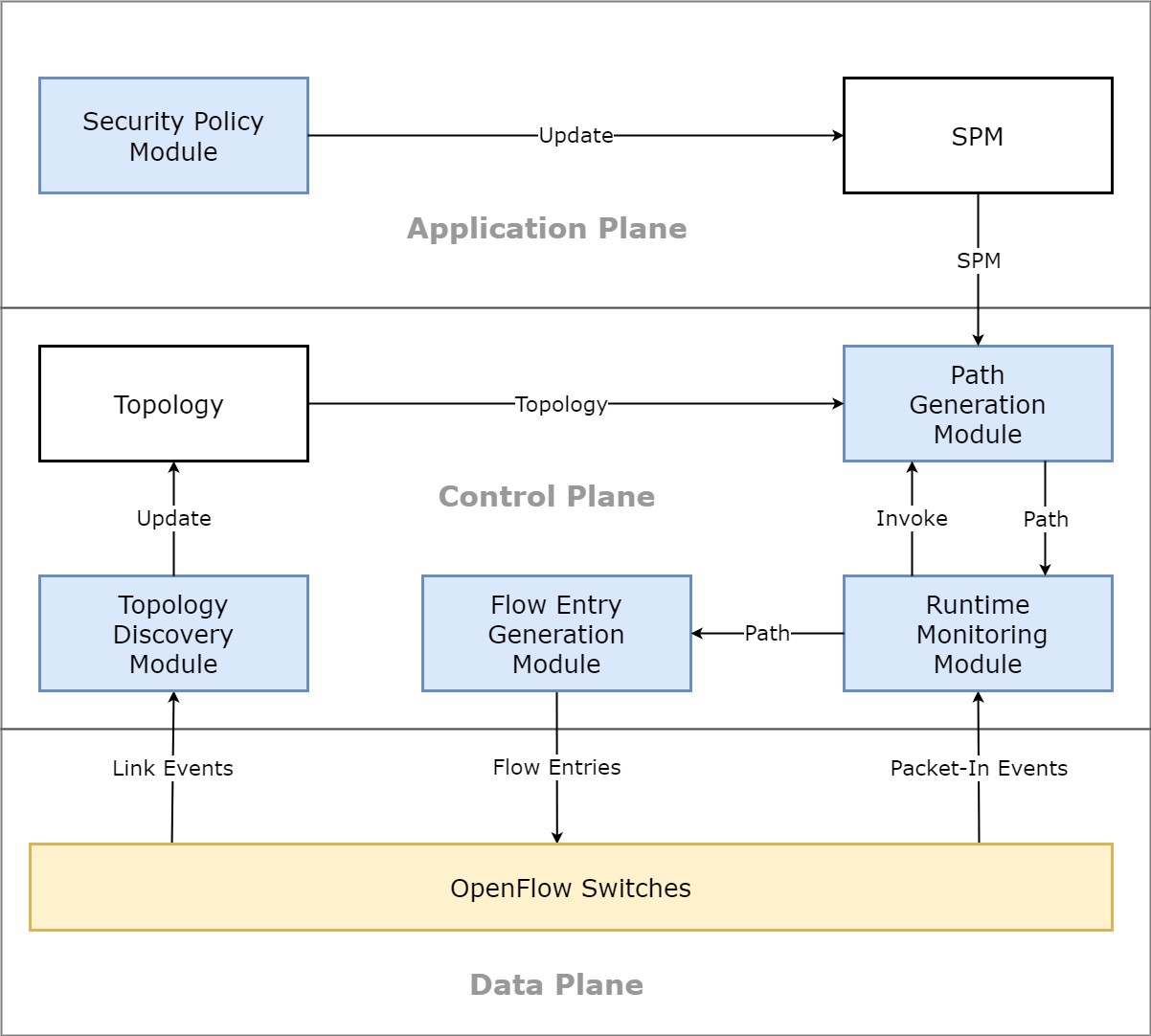}}
\caption{ The security policy transformation framework for SDN networks.}
\end{figure}

\subsection{Overview of the Framework}
$\bullet$ Security policy module is deployed in the application plane and responsible for maintaining the security policy (SPM). Each access control relationship $R_i\in$ SPM is designed as a 3-tuple: $($ $s_i$, $o_j$, $fixed$ $)$ based on the system model, where $s_i$ is the subject; $o_j$ is the object; $fixed$=$\{$ 0, 1 $\}$ is a tag bit, $fixed$=1 represents the relationship has been updated by user, $fixed$=0 represents it is unchanged. SPM is stored as a text document and can be updated by the user at runtime.

$\bullet$ Topology discovery module is deployed in the control plane and responsible for creating a dynamic real-time topology of the entire data plane by capturing the link events transmitted from the OpenFlow switches. Based on the system model, each edge in the topology is designed as a tuple: $e_i$ = $($ $sw_{src}$, $port_{src}$, $sw_{dst}$, $port_{dst}$, $using$, $c$ $)$, where $using$=$\{$ $True$, $False$ $\}$ is a tag bit and can be changed by the real-time link events at runtime, $using$=$True$ represents the edge can be used now, $using$=$False$ represents the edge is interrupted now. For the convenient of researching, the cost of each edge is set to 1, i.e., $c$ =1. The generated topology is stored as a text document and can also be updated by the real-time link events at runtime.

$\bullet$ Runtime monitoring module is deployed in the control plane and responsible for monitoring all the traffics in the data plane by capturing the real-time packet-in events transmitted from the OpenFlow switches. When a new packet-in event arrives in the controller, the module first invokes the path generation module to transform the latest security policy (SPM) into a set of connected paths in the data plane, then invokes the flow entry generation module to transform all the connected paths into their corresponding flow entries deployed in the OpenFlow switches which are passed by these paths. As SPM and the topology of data plane will be evolved with the runtime environment, this module is designed to be triggered by the real-time packet-in events continuously, so that, when SPM is changed (i.e., occurring $fixed$=1) or the topology is changed (i.e., occurring $using$=$False$) at runtime, the module will first delete all the current flow entries deployed in the OpenFlow switches, then update all the flow tables by using the latest generated flow entries, so as to ensure the security properties defined by SPM can be synchronously and continuously hold in the data plane at runtime.

$\bullet$ Path generation module is deployed in the control plane and invoked by the runtime monitoring module at runtime. The module is responsible for transforming each $R_i\in$ SPM input from the runtime monitoring module into a corresponding path $P_i$ in the data plane by using the latest topology file and the path searching algorithm. Specifically, it first transforms the subject $s_i\in R_i$ and the object $o_j\in R_i$ into the hosts $h(s_i)$ and $h(o_j)$ in the data plane respectively, then finds a shortest connected path $P_i$ between $h(s_i)$ and $h(o_j)$ by using the path searching algorithm, finally all the connected paths transformed from SPM are returned to the runtime monitoring module.

$\bullet$ Flow entry generation module is also deployed in the control plane and invoked by the runtime monitoring module at runtime. The module is responsible for transforming the connected path into a set of flow entries deployed in the OpenFlow switches which are passed by the path, then utilizing the instructions provided by the controller to generate the practical flow entries and distributing these flow entries to the corresponding OpenFlow switches at runtime.

\subsection{Runtime Monitoring Algorithm}
The runtime monitoring algorithm deployed in the runtime monitoring module plays the role of coordinator in the framework, and can be described as Algorithm 1 in pseudo code. First of all, the algorithm creates two dynamic lists $S$ and $T$ by reading the latest SPM and Topo file respectively. If there existing an access control relationship has been changed by user ($R_i$.$fixed$=1) or a edge has been shut down in the topology ($e_i$.$using$=$False$) at runtime, it will clear all the current flow entries deployed in the OpenFlow switches for ready of the updating. In the following, for each access control relationship $R_i\in S$, it maps the subject $s_i\in R_i$ and the object $o_j\in R_i$ with the switches $sw_{src}$ and $sw_{dst}$ in the data plane, then transforms $R_i$ into a corresponding connected path $P_i$ by invoking the path searching algorithm djk-route($sw_{src}$, $sw_{dst}$, $N$). Based on the transformation rules, $P_i$ can be further transformed into a set of flow entries. When all the relationships in the List $S$ having been transformed, SPM has been transformed into a corresponding set of flow entries $\Delta$, the algorithm invokes the flow entry generation module to update the data plane by using $\Delta$. Since the algorithm is designed to be triggered by the packet-in events at runtime, so that it ensures the framework can perceive any changes in time when the security policy or the topology of data plane has been evolved with the environment, and then update the data plane synchronously at runtime.

\subsection{Path Searching Algorithm}
Another important algorithm in the framework is the path searching algorithm deployed in the path generation module. The algorithm is improved from the classic Dijkstra algorithm and can be described as the Algorithm 2 in pseudo code. First of all, the algorithm creates a dynamic matrix $djk$[$N$][$N$] by using the sum of Openflow switches $N$ and the topology file. In the following, it calculates a shortest connected path between $sw_{src}$ and $sw_{dst}$ in the data plane by using $djk$[$N$][$N$] and the created stacks. After multi-round calculating, the shortest path $P_i$ between $sw_{src}$ and $sw_{dst}$ is found and returned to the runtime monitoring module. As the cost of each edge has been set to 1 and not considering of the quality of services (QoS) of edges, so that the shortest path $P_i$ found by the algorithm is generated by calculating the minimum number of hops in the topology. Moreover, since the matrix $djk$[$N$][$N$] is dynamically created by the topology file, so that the searched shortest path will be evolved with the changing of the topology at runtime.

\section{Implementation and Evaluations}
In order to validate the feasibility and effectiveness of the framework proposed in Section 4, we set up an experimental system and implement the framework with POX controller\cite{POX} and Mininet emulator\cite{Mininet}. First of all, we implement a virtual SDN network by using the Mininet emulator. As shown in Figure 3, the topology of the network consists of 6 hosts ($h_1\sim h_6$) and 11 OpenFlow switches ($sw_{1}\sim sw_{11}$). We further implement the security policy module, topology discovery module, runtime monitoring module, path generation module and flow entry generation module with Python 3.6.1 and integrate these modules with the core of POX controller. The experimental system consists of a Lenovo workstation with Windows OS, Intel-i7 32Cores 2.60GHz CPU, 32GB RAM and a Raspberry platform with Linux OS, ARM-v7 CPU and 945MB RAM. The POX controller and the functional modules are deployed in Lenovo workstation, Mininet emulator is deployed in Raspberry platform, and Lenovo workstation is connected with Raspberry platform using coaxial cable directly.
\begin{table}[htbp]
\caption{The high-level security policy (SPM)}
\begin{center}
\begin{tabular}{c|c}
\hline
$R_{1}$ &
$($ $1$, $5$, $1$ $)$\\
\hline
$R_{2}$ &
$($ $5$, $1$, $1$ $)$\\
\hline
$R_{3}$ &
$($ $2$, $4$, $1$ $)$\\
\hline
$R_{4}$ &
$($ $4$, $2$, $1$ $)$\\
\hline
\end{tabular}
\end{center}
\end{table}

\subsection{Effectiveness Evaluation }
The security policy used for validating the effectiveness of the framework is shown in Table I. Since any effective interaction is bidirectional in SDN networks, i.e., the subject's host and the object's host must be ensured they can access each other in the data plane, so that we design the security policy as 4 access control relationships ($R_1\sim R_4$) to ensure $h_1$ (1) and $h_5$ (5) can access each other, $h_2$ (2) and $h_4$ (4) can access each other, and all the relationships of SPM are set as $fixed$=1, i.e., having been updated by user. In the following, the effectiveness evaluations towards the framework will be carried out under 4 different scenarios at runtime, they are the effectiveness after loading the flow entries, the effectiveness after cutting down the path and the effectiveness after changing SPM.
\begin{figure}[htbp]
\centering
\scalebox{0.22}{\includegraphics{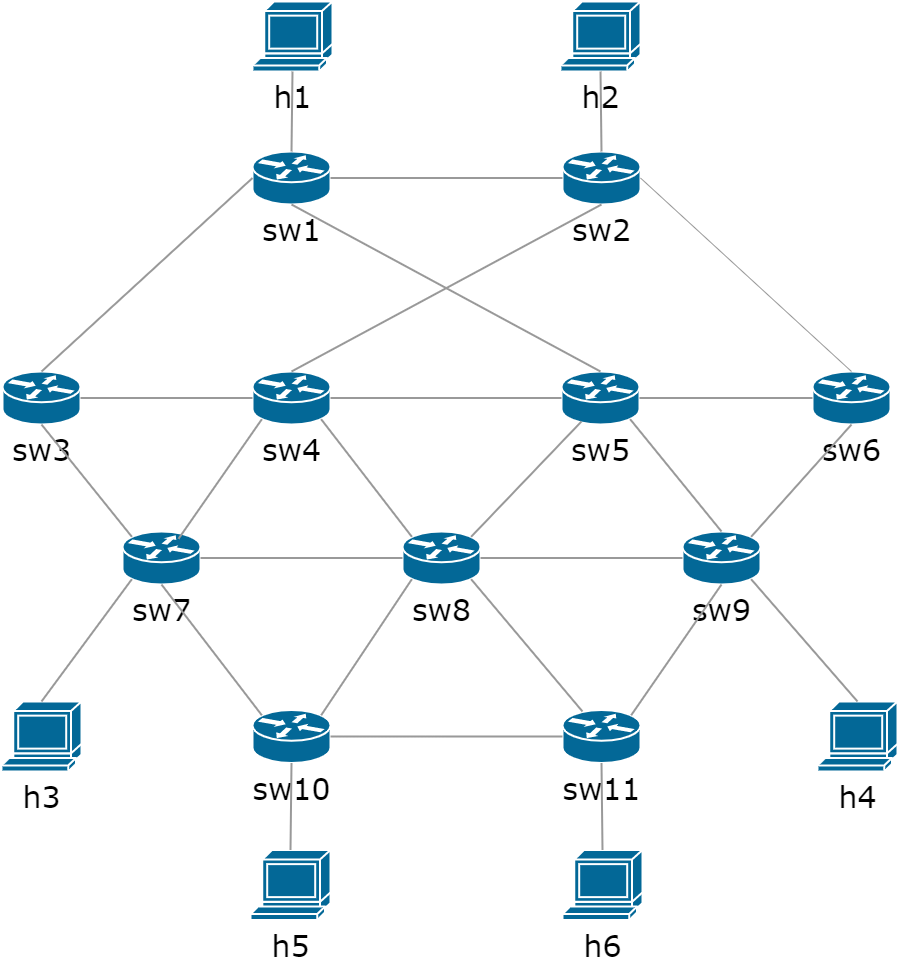}}
\caption{ The topology of the virtual SDN network.}
\end{figure}

\subsubsection{Effectiveness after Loading the Flow Entries}
The purpose of this experiment is to validate whether the data plane after loading the generated flow entries can hold the security properties defined by SPM. First of all, the subjects and objects of SPM shown in Table I, i,e., the $1$, $2$, $4$ and $5$, are transformed into their corresponding hosts in the data plane by using security policy transformation. Specifically, the $1$ is transformed into $h_1$=$($ 10.0.0.1, $sw_1$, 1 $)$, the $2$ is transformed into $h_2$=$($ 10.0.0.2, $sw_2$, 1 $)$, the $4$ is transformed into $h_4$=$($ 10.0.0.4, $sw_9$, 1 $)$ and the $5$ is transformed into $h_5$=$($ 10.0.0.5, $sw_{10}$, 1 $)$ respectively. In the following, the path searching algorithm, i.e., Algorithm 2, searches the shortest path between the subject's host and the object's host for each $R_i\in$ SPM  based on the latest topology generated from the topology file.
\begin{figure}[htbp]
\centering
\scalebox{0.22}{\includegraphics{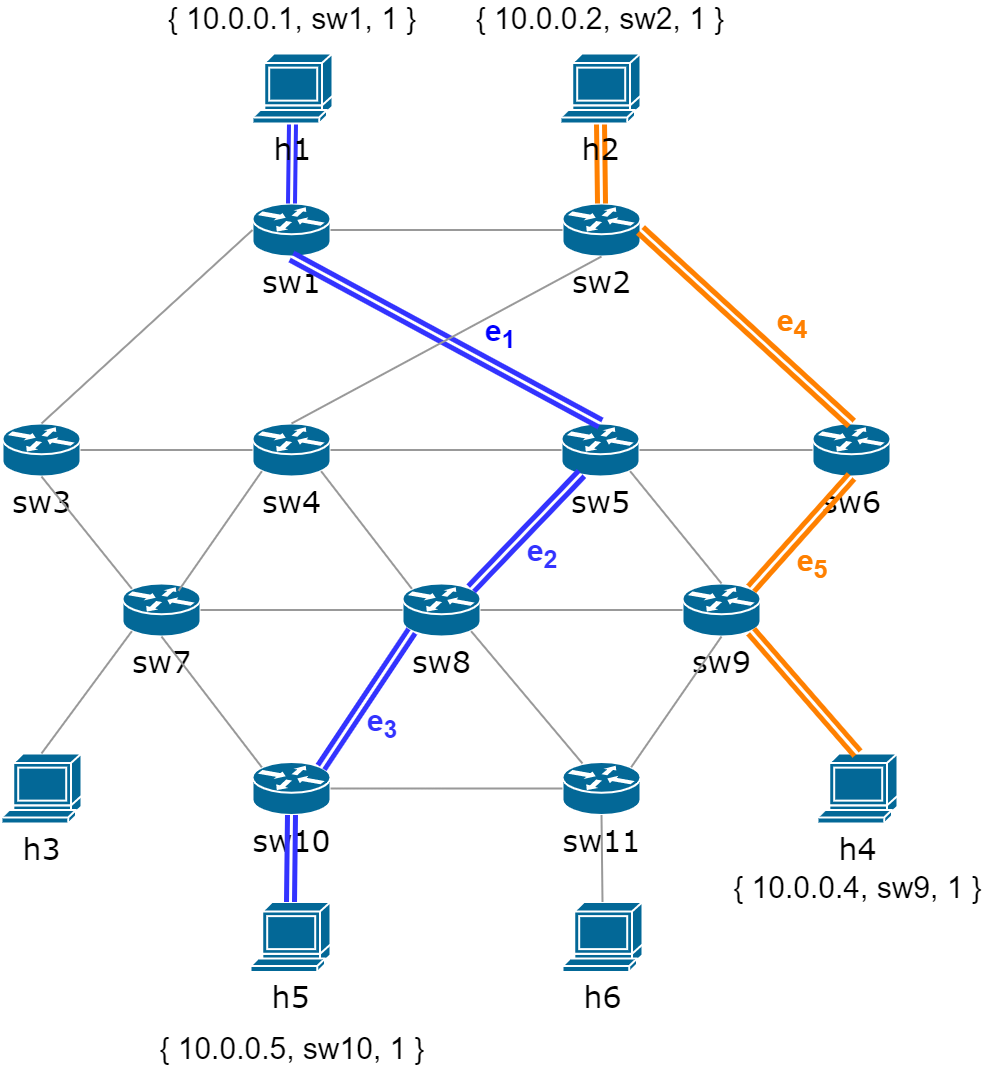}}
\caption{ The shortest paths searched by Algorithm 2.}
\end{figure}

After that, the relationships shown in Table I have been transformed into 4 corresponding shortest connected paths ($P_1\sim P_4$) in the data plane. As shown in Figure 4, the $R_1$ is transformed into $P_1$=$\{$ $h_1$, $e_1$, $e_2$, $e_3$, $h_5$ $\}$, the $R_2$ is transformed into $P_2$=$\{$ $h_5$, $e_3$, $e_2$, $e_1$, $h_1$ $\}$, the $R_3$ is transformed into $P_3$=$\{$ $h_2$, $e_4$, $e_5$, $h_4$ $\}$ and the $R_4$ is transformed into $P_4$=$\{$ $h_4$, $e_5$, $e_4$, $h_2$ $\}$, where the $P_1$ and $P_2$ are depicted with the blue lines, the $P_3$ and $P_4$ are depicted with the orange lines. In the following, the flow entry generation module transforms each path into a set of flow entries deployed in the switches passed by the path. Specifically, $P_1$ and $P_2$ are transformed into the flow entries deployed in the switches $\{$ $sw_{1}$, $sw_5$, $sw_8$, $sw_{10}$ $\}$, while $P_3$ and $P_4$ are transformed into the flow entries deployed in the switches $\{$ $sw_2$, $sw_6$, $sw_9$ $\}$.
\begin{figure}[htbp]
\centering
\scalebox{0.45}{\includegraphics{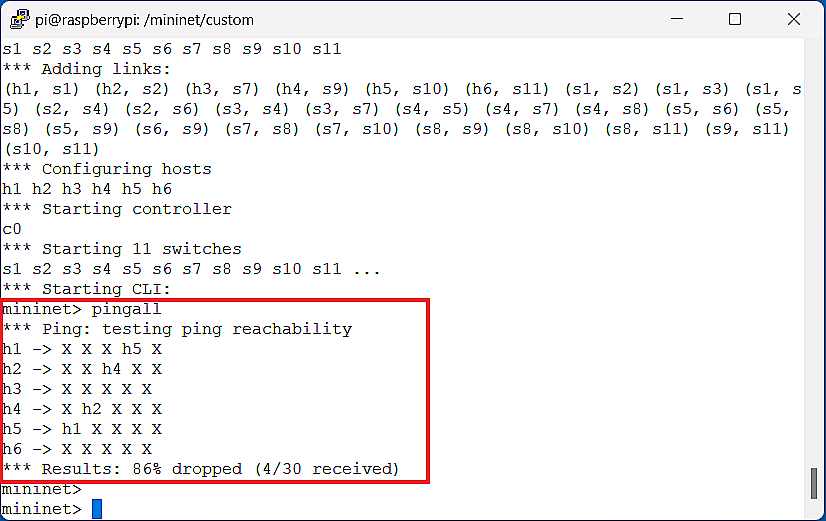}}
\caption{ The result of executing the $pingall$ instruction.}
\end{figure}

After that, we execute the $pingall$ instruction in the Mininet CLI and observe the reachability of the entire data plane. As shown in Figure 5, $h_1$ and $h_5$ can access each other, $h_2$ and $h_4$ can access each other either, so that the data plane after loading the generated flow entries has been proven that it can hold all the security properties defined by SPM.

\subsubsection{Effectiveness after Cutting down the Path}
The purpose of this experiment is to validate whether the security properties defined by SPM can be continuously hold in the data plane after the connected path between the subject's host and the object's host is shut down at runtime. If the framework can synchronously perceive this change from the topology and automatically find another new connected path to keep the data plane holding the security properties at runtime, then the framework will be proven to be effective under this scenario. First of all, we let $h_1$ and $h_5$ can access each other by loading the corresponding flow entries into the data plane, and the shortest path between $h_1$ and $h_5$ is initialed with $\{$ $h_1$, $e_1$, $e_2$, $e_3$, $h_5$ $\}$. In the following, we make a continuous TCP traffic sent from $h_1$ to $h_5$ by using the $iperf$ instruction in the Mininet CLI, we set the duration time of the experiment equals 60 seconds and record the throughput of the traffics in $h_5$. When the time reaches 23 seconds, we shut down the edge between $sw_8$ and $sw_{10}$ existed in the path $P_1$ by using the instruction in the Mininet CLI and let the experiment going on. When the experiment is finished, we read the data recorded in $h_5$ and plot them in Figure 6. As shown in Figure 6, the throughput of the traffic sent from $h_1$ to $h_5$ is sharply declined after the edge is shut down at 23 seconds, and completely becomes zero from 25 seconds to 37 seconds. After 38 seconds, the traffic quickly returns to normal until the end of the experiment. The experimental result has illustrated that the framework can synchronously perceive the change caused by cutting down one edge between $sw_8$ and $sw_{10}$, and automatically find another new shortest connected path between $h_1$ and $h_5$, i.e., $\{$ $h_1$, $e_6$, $e_7$, $e_8$, $h_5$ $\}$, so as to make the traffic returning to normal quickly and keep the data plane holding the security properties defined by SPM at runtime. The new shortest connected path searched by Algorithm 2 after cutting down the using path is shown in Figure 7.
\begin{figure}[htbp]
\centering
\scalebox{0.35}{\includegraphics{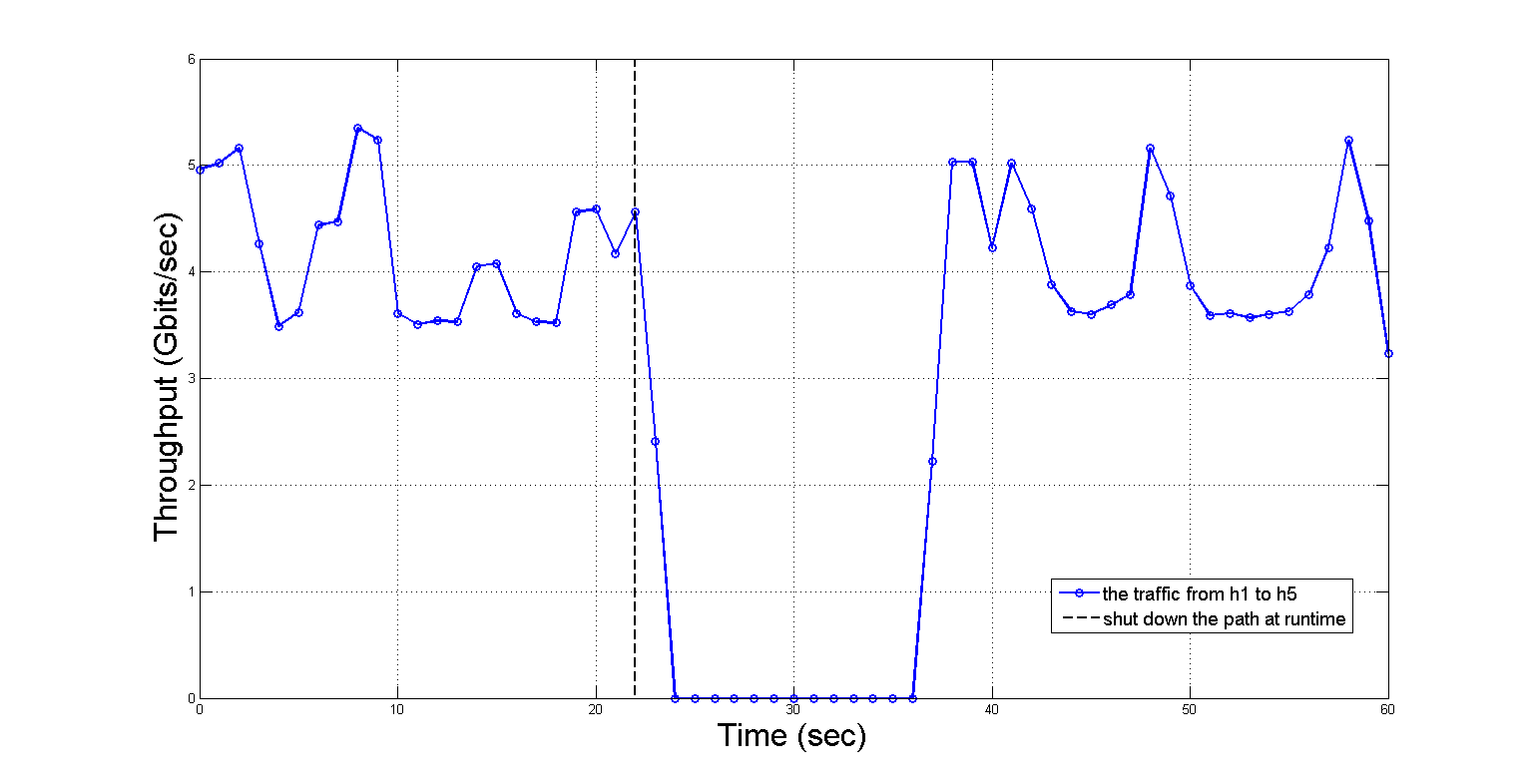}}
\caption{ The blue line represents the throughput of the traffic sent from $h_1$ to $h_5$. The dotted line represents we shut down the connected path at 23 seconds.}
\end{figure}

\begin{figure}[htbp]
\centering
\scalebox{0.22}{\includegraphics{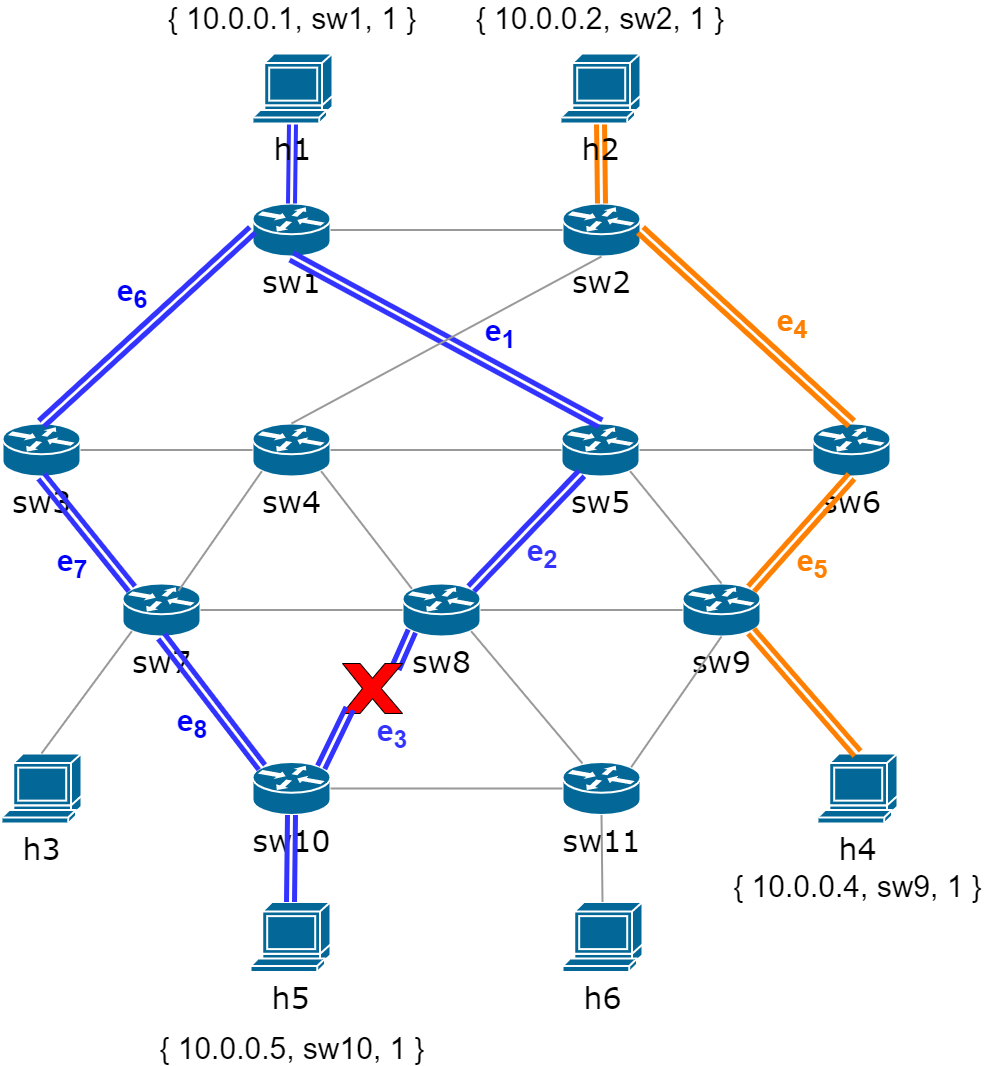}}
\caption{ The new connected path searched by Algorithm 2 at runtime.}
\end{figure}

\subsubsection{Effectiveness after Changing SPM}
The purpose of this experiment is to validate whether the security properties defined by SPM can be continuously hold in the data plane after SPM is changed by the user at runtime. If the framework can synchronously perceive this change from SPM and automatically update the flow entries deployed in the switches to keep the data plane holding the evolved security properties at runtime, then the framework will be proven to be effective under this scenario. First of all, we let $h_1$ and $h_5$ can access each other by loading a corresponding flow entries into the data plane, then make a continuous TCP traffic sent from $h_1$ to $h_5$ by using the $iperf$ instruction in the Mininet CLI, we set the duration time of the experiment equals 60 seconds and record all the throughput data of the traffic in $h_5$.

In the following, we first validate the effectiveness of the framework under the scenario of changing SPM from $h_1$ and $h_5$ can access each other to $h_2$ and $h_5$ can access each other at runtime, and the experimental result under this scenario is plotted in Figure 8. As shown in Figure 8, the throughput of the traffic sent from $h_1$ to $h_5$, which is depicted with the red line, is quickly declined when we load the new SPM into the controller at 29 seconds, and completely becomes zero after 31 seconds. From 31 seconds until to the end of the experiment, $h_5$ can only receive the traffic sent from $h_2$ which is depicted with the blue line. The experimental result illustrates the framework can synchronously perceive this change and keep the data plane holding the evolved security properties after changing SPM from $h_1$ and $h_5$ can access each other to $h_2$ and $h_5$ can access each other at runtime.
\begin{figure}[htbp]
\centering
\scalebox{0.35}{\includegraphics{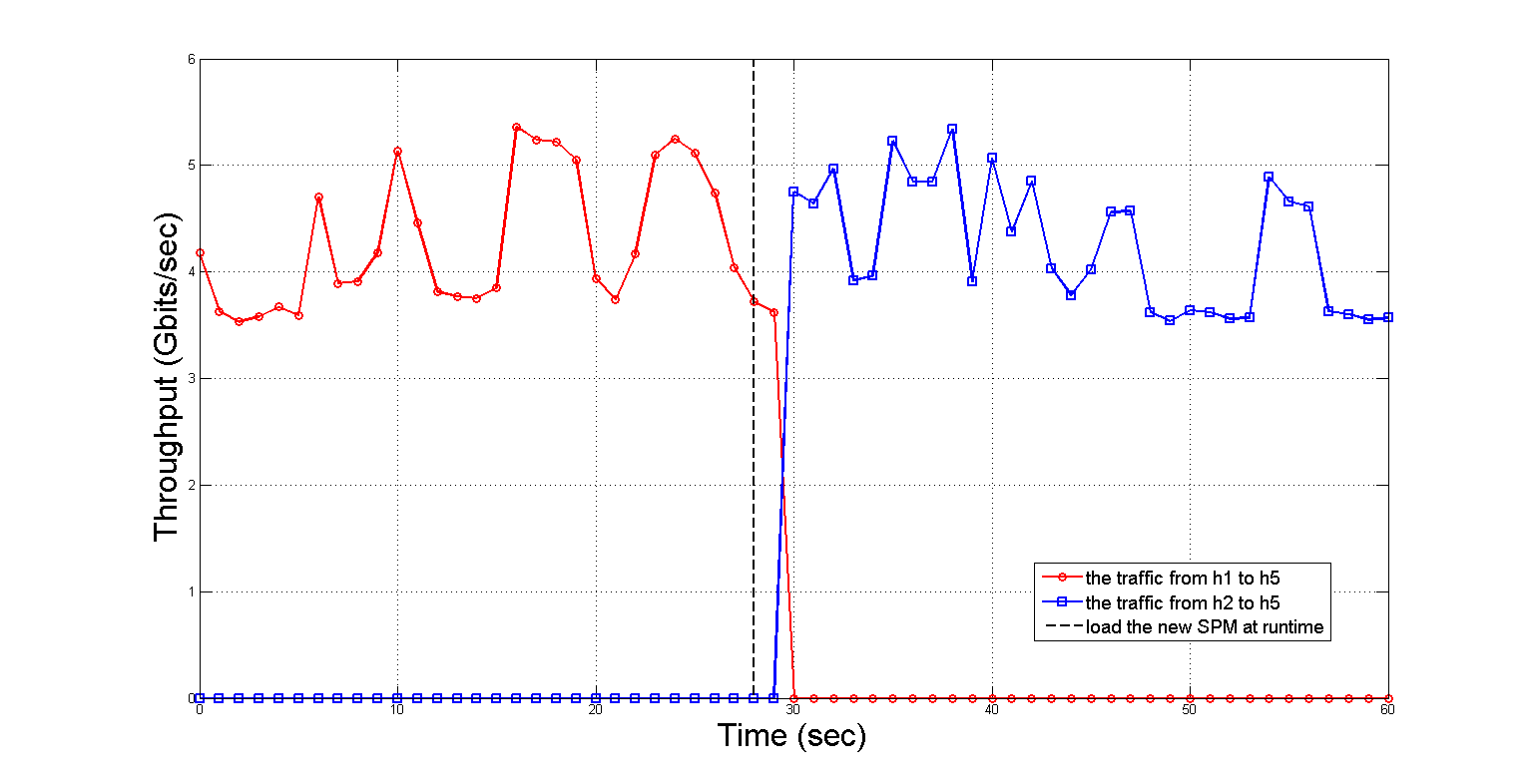}}
\caption{ The red line represents the throughput of the traffic sent from $h_1$ to $h_5$, the blue line represents the throughput of the traffic sent from $h_2$ to $h_5$, the dotted line represents we change SPM at 29 seconds.}
\end{figure}

After that, we further validate the effectiveness of the framework under the scenario of adding a new relationship into SPM, i.e., adding $h_2$ and $h_5$ can access each other, at runtime. The experimental result under this scenario is plotted in Figure 9. As shown in Figure 9, the throughput of the traffic sent from $h_1$ to $h_5$, which is depicted with the red line, still keeps normal before we load the new SPM into the controller at 43 seconds. From 44 seconds until to the end of the experiment, $h_5$ can receive the continuous traffic sent from $h_2$ which is depicted with the blue line, and can also receive the traffic sent from $h_1$ at the same time. Due to the crowding of the traffic sent from $h_2$, the throughput of the traffic from $h_1$ is declined from 3.71GB/s to 1.99 GB/s. The throughput of the traffic sent from $h_2$ is still kept between 1.6GB/s and 2.2GB/s after 44 seconds. The experimental result illustrates the framework can synchronously perceive this change and keep the data plane holding the evolved security properties after adding a new relationship into SPM at runtime.
\begin{figure}[htbp]
\centering
\scalebox{0.35}{\includegraphics{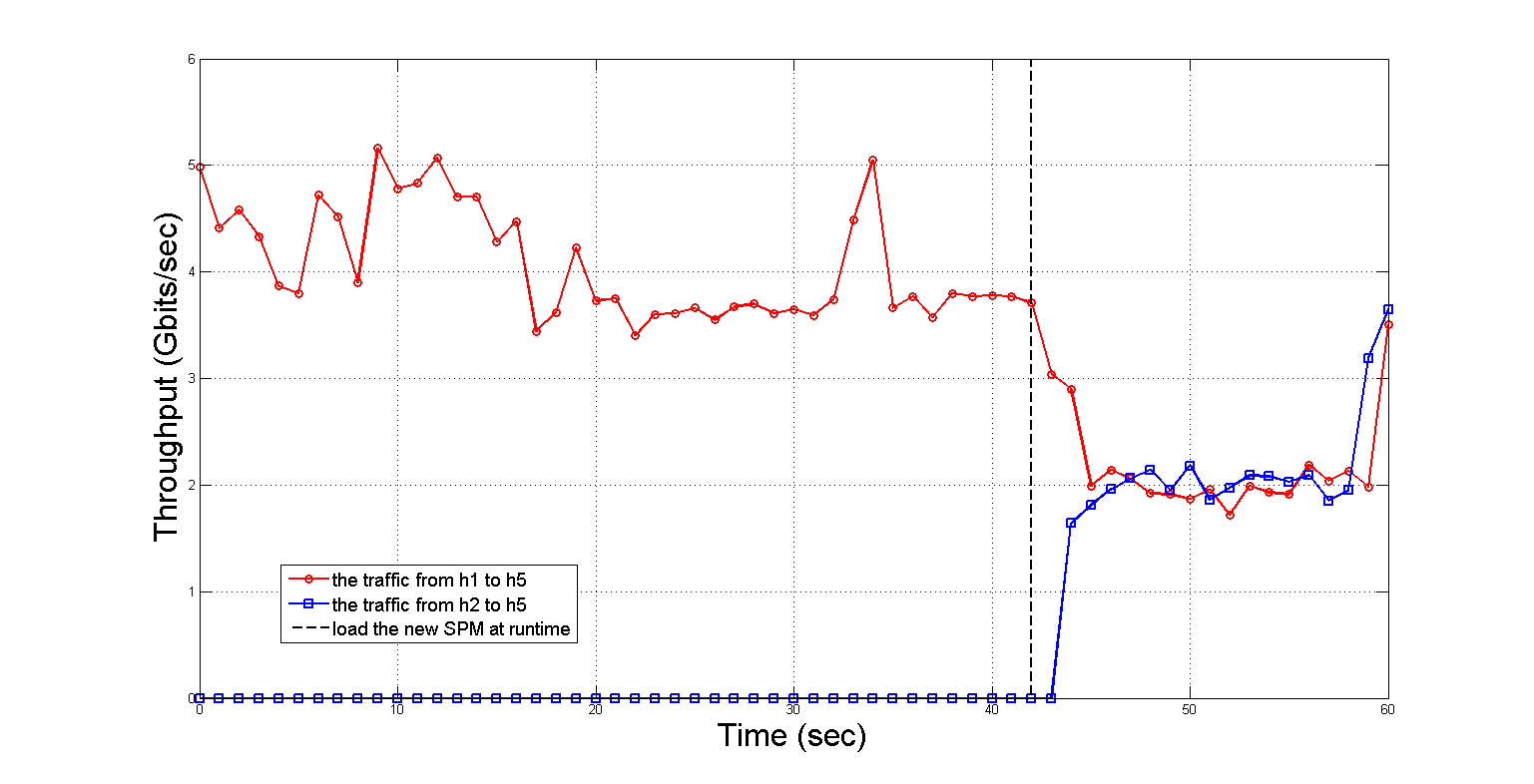}}
\caption{ The red line represents the throughput of the traffic sent from $h_1$ to $h_5$, the blue line represents the throughput of the traffic sent from $h_2$ to $h_5$, the dotted line represents we change SPM at 43 seconds.}
\end{figure}

\subsection{Performance Evaluation}
As the critical algorithm used for implementing the security policy transformation, the performance of the path searching algorithm, i.e., Algorithm 2, needs to be further evaluated. First of all, the sum of access control relationships of the security policy (SPM) is denoted as $M$, and the sum of OpenFlow switches in the topology is denoted as $N$ in this performance evaluation. Then by leveraging the Python programming, the execution time of Algorithm 2 have been recorded in milliseconds (ms) for calculating the shortest paths under setting the different value of $M$ and $N$. The experimental result is plotted in Figure 10. As shown in Figure 10, with gradually amplifying the value of $M$ from 2 to 10, and the value of $N$ from 11 to 400 respectively, the execution time of Algorithm 2 shows an obvious exponential upward trend. Moreover, according to the description of Algorithm 2, the time complexity for calculating only one shortest path will reach $\mathcal{O}$($N^{2}$), because the Algorithm 2 needs to create a dynamic matrix $djk[N][N]$ and further calculates the while loop, so that the time complexity for transforming all the access control relationships defined by SPM into their corresponding shortest paths will reach $\mathcal{O}$($M\times N^{2}$).
\begin{figure}[htbp]
\centering
\scalebox{0.35}{\includegraphics{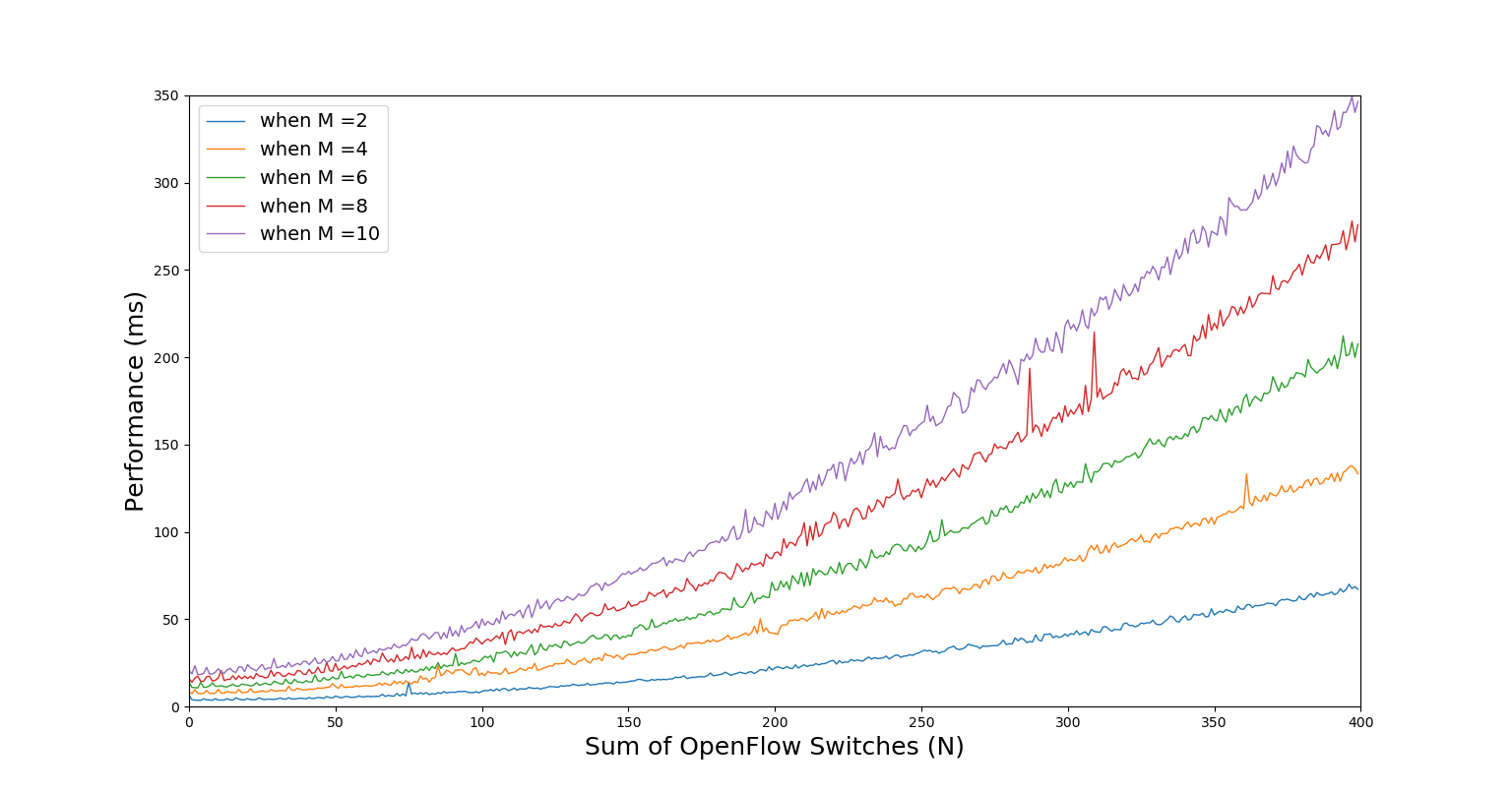}}
\caption{ The execution time of Algorithm 2 recorded in milliseconds (ms) for calculating the shortest paths under setting the different value of $M$ and $N$, where $M$ represents the sum of access control relationships of SPM, $N$ represents the sum of OpenFlow switches in the topology. }
\end{figure}

\section{Conclusion}
In this paper, we propose a practical runtime security policy transformation framework for SDN networks. First of all, we specify the security policies used by SDN networks, such as access control policies or firewall policies, as a system model of security policy (SPM). SPM is of a high-level system model without containing any underlying information of data plane. From the theoretical level, we establish the system model for SDN network and propose a formal method to transform SPM into the corresponding flow entries automatically. The flow entry transformed from SPM is of a low-level system model containing the underlying information of data plane. From the practical level, we propose a runtime security policy transformation framework which consists of the security policy module, topology discovery module, runtime monitoring module, path generation module, as well as flow entry generation module. Leveraging these functional modules, the framework can solve the problem of how to find a connected path for each relationship defined by SPM in the data plane, how to transform the path into the system model of flow entries, as well as how to generate the practical flow entries by using the system model of flow entries. In order to validate the feasibility and effectiveness of the framework, we set up an experimental system and implement the framework by using POX controller and Mininet emulator. The experimental result illustrate the framework is completely effective at runtime.

However, there still exists some problems needed to be further researched in the future. The current path searching algorithm, i.e., Algorithm 2, used by the framework is improved from the classic Dijkstra algorithm and finds the shortest path by calculating the minimum number of hops in the topology. However, in the real SDN networks, the problem of searching a connected path between the two hosts need to consider the matters of quality of service (QoS), load balance, as well as some specific requirements about the traffic engineering at runtime, so that the framework needs to be further improved by employing some novel path searching algorithms based on multi-object optimization\cite{optimization} or reinforcement learning\cite{reinforcement} methods.

\section*{Acknowledgment}
This paper has been sponsored and supported by National Key Research and Development Program of China (Grant No.2018YFB0803400),
Doctoral Foundation of Qingdao Binhai University (Grant No. BS2022A10), partially supported by Key Program of National Natural Science Foundation of China (Grant No.61932013).

\section*{References}
\bibliographystyle{elsarticle-num}
\bibliography{2023-ELS-02}


\end{document}